\documentclass[aps,prl,twocolumn,showpacs,groupedaddress,amsmath,amssymb]{revtex4}
\usepackage{graphicx}
\usepackage{color}
\usepackage{bm}

\begin{document}

\title{Non-Hermiticity Induced Flat Band}
\author{ Hamidreza Ramezani$^{1}$} 
\email {hamidreza.ramezani@utrgv.edu}
\affiliation{$^1$Department of Physics, The University of Texas Rio Grande Valley, Brownsville, TX 78520, USA}

\begin{abstract}
We demonstrate the emergence of an entire flat band with no complex component embedded in dispersive bands at the exceptional point of a PT symmetric photonic lattice. For this to occur, the gain and loss parameter effectively alters the size of the partial flat band windows and band gap of the photonic lattice simultaneously. The mode associated with the entire flat band is robust against changes in the system size and survives even at the edge of the lattice. Our proposal offers a route for controllable localization of light in non-Hermitian systems and a technique for measuring non-Hermiticity via localization.
\end{abstract}


\maketitle
Controllable and yet robust confinement of light , or simply localization, is vital for many applications such as quantum simulation of nondispersive states, diffraction-less long distances light propagation, enhancement of nonlinear effects, stop light, and imaging. Various approaches have been proposed to achieve localization, among them are impurity in periodic systems \cite{1, 2, 3}, quasi-periodic systems \cite{t1,t2,t3}, Anderson localization \cite{4}, nonlinear self-trapping \cite{5, t4}, bound state in continuum (BIC) \cite{6,7}, and flat bands \cite{8,9}. In flat bands, localization occurs due to a destructive interference of the geometric phases and observed recently in Lieb photonic lattices \cite{8,9}. Interest in flat bands is not limited to optics \cite{10,11} or photonics \cite{8,9,12,13} and has been studied in graphene \cite{14}, superconductors \cite{15,16}, quantum Hall effect \cite{17,18,19,20}, and exciton polariton condensates \cite{21,22,23}.

Nevertheless, all these achievements are limited in view of studying the properties of flat bands using Hermitian potential \cite{konotop}. Thus, many of these studies cannot be used in active systems, such as coupled laser cavities and metal-dielectric structures, where gain or loss exists naturally. Moreover, none of the aforementioned studies can control the generated flat band and the localized modes. In other words, entering to the delocalization regime on demand and at different propagation coupling length/time needs fabrication of different samples. Overcoming these limitations will not only enrich the conventional research in flat bands, but also offer new methods for controllable localization and imaging technologies. It is therefore extremely desirable to investigate and propose architectures with flat band that incorporate gain and loss mechanisms.
\begin{figure}
	\includegraphics[width=1\linewidth, angle=0]{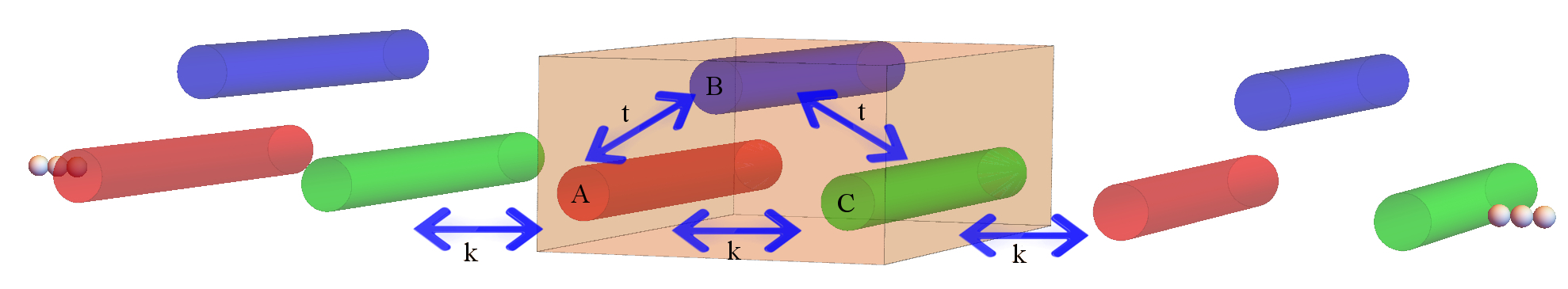}
	\caption{(Color online) Schematic of a quasi 1D PT symmetric array of coupled waveguides with a flat band at the exceptional point. The unit cell of the array (identified in a cuboid) composed of a gain waveguide (\textit{A}), a passive waveguide (\textit{B}), and a loss waveguide (\textit{C}). The passive waveguide in the unit cell is coupled to the gain and loss waveguides with coupling strength $t$. The gain and loss waveguides are coupled with coupling $k$.\label{fig1}}
\end{figure}
\begin{figure}
	\includegraphics[width=1\linewidth, angle=0]{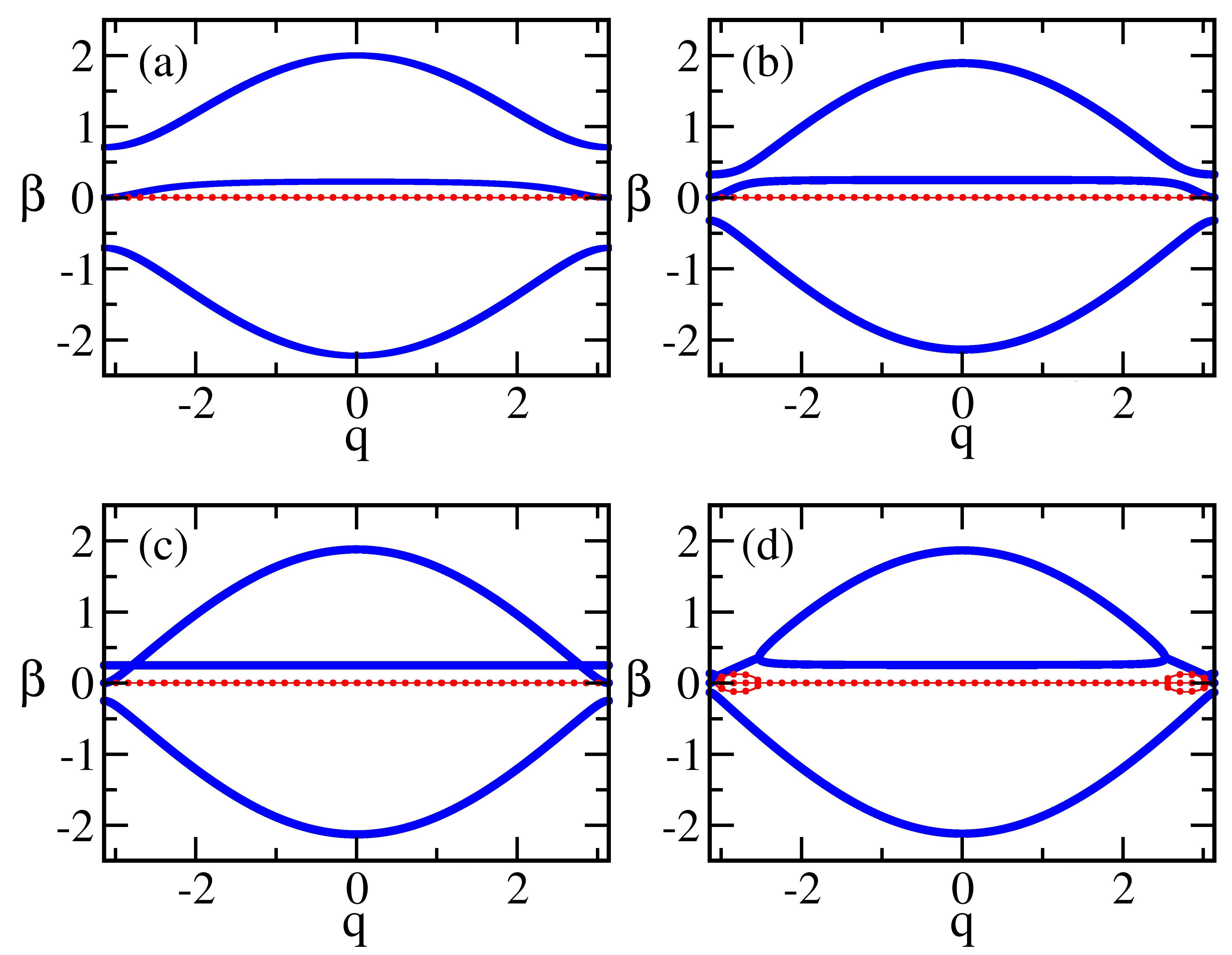}
	\caption{(Color online)  Real (blue curves) and imaginary (red dots) part of the propagation constant $\beta$ of the PT symmetric structure in Fig. \ref{fig1} vs the Bloch wavevector $q$ with $k=1$ and $t=0.5$. (a) Hermitian case with $\gamma=0$. The imaginary parts of the propagation constants are zero. Three bands are separated by gaps. The bands are partially flat at the edge and center of Brillouin zone. (b) Exact phase with $\gamma=0.95\gamma_{PT}$. The imaginary parts of the propagation constants are zero. By increasing the $\gamma$ the bands approach each other, namely the gaps become smaller. For the two upper bands the window for which the bands are partially flat expands. (c) Exceptional point with $\gamma=\gamma_{PT}\approx 0.66$. The two upper bands touch each other at $q\approx2.77$, the partial flat bands of each one combine and form an entire flat band. Still imaginary parts are zero. The flat band is embedded in between the two dispersive bands and forming an infinite number of BIC-like states. (d) Broken phase with $\gamma=1.05\gamma_{PT}$. The two upper band start merging for $\gamma>\gamma_{PT}$. The flat band lose its flatness at the windows of wavevector for which the bands are merged. The propagation constant of the merged parts becomes complex (red dots with non-zero values). \label{fig2}}
\end{figure}

Here we show that by altering the degree of non-Hermiticity in a PT symmetric lattice one can control the flat band formation and effectively control the localization of light. More specifically, we show that an entire flat band can be obtained at the exceptional point of a PT symmetric lattice which results in non-diffracting beam propagation with constant intensity. Below (above) the exceptional point, as we reduce (increase) the gain and loss parameter, flat band shrinks and form partial flat bands. The exceptional point induced flat band is located, with no gap, in between the dispersive bands and therefore composed of infinite BIC-like states in complex lattices. While in the exact phase partial flat bands are separated from each other with a gap and therefore are not BIC-like states, in the broken phase the partial flat bands are in between the dispersive bands with no gap and thus can be considered as BIC-like states. Furthermore, we show that localized state associated with the bulk is robust against system size and remains localized at the edge of the lattice. Our results provide a scheme for generation of controllable flat bands and BIC-states in synthetic non-Hermitian lattices.

To show how exceptional point can induce an entire flat band, we consider a quasi 1D PT symmetric waveguide array depicted schematically in Fig.\ref{fig1}. The unit cell of the waveguide array is a tri-mer (indicated with a box in Fig.\ref{fig1}) that consists of a gain waveguide (\textit{A}) with gain parameter $\gamma$, a passive waveguide (\textit{B}) with no gain or loss, and a loss waveguide (\textit{C}) with loss parameter $-\gamma$. Each of the waveguides supports only one mode. In the \textit{A-B-C} units, the gain and loss waveguides are evanescently coupled to the passive waveguide with coupling strength $t$. In our model gain (loss) waveguides are coupled to nearest neighbor loss (gain) waveguides with coupling strength $k>t$. Interestingly, it has been shown that the unit cell of our array can act as a unidirectional laser/absorber \cite{28,29}. 

With a very good approximation diffraction of the electric field amplitude at the nth unit cell $\Psi_n=(A_n,B_n,C_n)^T$ is given by 
\begin{equation}
\begin{array}{c}
i\frac{dA_n}{dz}=i\gamma A_n-tB_n-k(C_n+C_{n-1})\\
i\frac{dB_n}{dz}=-t(A_n+C_n)\\
i\frac{dC_n}{dz}=-i\gamma C_n-tB_n-k(A_n+A_{n+1})
\end{array}
\label{eq0}
\end{equation}
Above $z$ is the propagation direction in the unit of couplings and we assumed that the real part of the refractive index of all the waveguides are zero. A non-zero real part shifts the whole dispersion relation. Adopting momentum representation in Eq.(\ref{eq0}) $\Psi_n=\frac{1}{2\pi}\int_{-\pi}^{\pi}dq\psi_q e^{inq}$, with $\psi_q=(a,b,c)^T$, results in the following Schr\"{o}dinger equation for each value of $q$ 
\begin{equation}
i\dot\psi_q=H_q\psi; H_q=\left(\begin{array}{ccc}
i\gamma&-t&-k-ke^{-iq}\\
-t&0&-t\\
-k-ke^{iq}&-t&-i\gamma
\end{array}\right)
\label{eq1}
\end{equation}
Where $\cdot{}$ is derivative with respect to $z$.

In Figure \ref{fig2}, we plotted the dispersion relation of the $H_q$ for different values of the gain and loss parameter. Specifically, Figs. \ref{fig2}a-d depict the dispersion for the Hermitian case with $\gamma=0$, exact phase with $\gamma<\gamma_{PT}=0.95\gamma_{PT}$, exceptional point with $\gamma=\gamma_{PT}$, and broken phase with $\gamma=1.05\gamma_{PT}$. In the Hermitian case we observe that the dispersion relation of the lattice has three bands separated by two gaps. All the bands are partially flat at the center and edges of the Brillouin zone, which is expected for Hermitian lattices. Figure \ref{fig2}b shows that by increasing $\gamma$ the bands come close to each other and the gaps become tighter. At the same time, the flat band windows of the middle band, at the center of Brillouin zone, and the upper band, at the edges of the Brillouin zone, become wider. As depicted in Fig \ref{fig2}c, at the exceptional point the upper and lower bands touch each other at $q\approx\pm2.77$. Thus, the two partial flat bands created by non-Hermiticity combine and form an entire flat band embedded between the rest of their dispersive bands. This entire flat band has a zero group velocity and infinite effective mass which is a new BIC-like state. By definition modes in dispersive bands should disperse. However, sometimes by means of symmetries a mode (BIC mode) might not diffract, although it belongs to the dispersive band. Thus a flat band in between two dispersive bands with no gap is a BIC state \cite{8,9}.

Eigenmodes of PT symmetric systems are bi-orthogonal which makes the total norm nonconservative. Furthermore, at any exceptional point of a non-Hermitian system at least two eigenvalues and eigenvectors coalesce and become degenerate \cite{24,30,t5} which makes the corresponding Hamiltonian to be defective \cite{32}. Thus, while it is excepted that an entire flat band of a Hermitian system allows the formation of compacton states with no diffraction \cite{31}, it is not obvious that our proposed flat band at the exceptional point supports non-dispersive modes with constant intensity. In the following we show that although at the exceptional point the Hilbert space of our system collapses, the system supports localized non-dispersive modes with constant intensity. In the broken phase with $\gamma>\gamma_{PT}$, (Fig \ref{fig2}d), the parts of the bands that are merged, lose their flatness while the unmerged part of the middle band remains flat. This partial flat band, which shrinks as we increase the gain and loss parameter, is in between the dispersive bands and the modes associated with it are BIC-like states. This is in contrast to previous studies \cite{24, 25,26}. Specifically, while the lattice supports a partially flat band, the merged parts of the bands are not flat. It should be noted that for very large values of gain and loss parameter and deep in the broken phase ultimately all the bands merge and form a flat band \cite{24, 25,26}. Such flat bands have eigenmodes with entirely degenerate real and asymmetric nondegenerate imaginary components. Consequently, the modes associated with these bands will amplify or decay exponentially in a non-uniform fashion \cite{27} which makes them less appealing for applications.

In our lattice, obtaining the entire flat band can be a measure for which the system reaches to the exceptional point. Assuming that the propagation constant of the lattice at the flat band is denoted by $\beta=\beta_0$, we can find the corresponding non-dispersive mode. In the momentum space and using Eq.(\ref{eq1}), we can calculate the dispersion-less mode associated with wavevector $q$. Indeed, one can find that this mode is given by $ \psi_q=\left(
\begin{array}{ccc}
1&-\frac{t}{\beta_0}(1+\xi)&\xi
\end{array}
\right)^T$ 
or 
$ \psi_q=\left(
\begin{array}{ccc}
1&-\frac{t}{\beta_0}(1+1/\xi^\ast)&1/{\xi^ \ast}
\end{array}
\right)^T$, 
where $\xi=(\beta_0-\frac{t^2}{\beta_0}-i\gamma)(\frac{t^2}{\beta_0}-k(1+e^{-iq}))^{-1}$ and $\ast$ means complex conjugation. As both expressions at the exceptional point denote the same mode they should be equal. Therefore, one can find the gain and loss value for which we attain the exceptional point, $\gamma_{PT}=\frac{\sqrt{2 \beta_0  \left(k^2+t^2\right)+2 k \cos (q) \left(k \beta_0 -t^2\right)-2 k t^2-\beta_0 ^3}}{\sqrt{\beta_0 }}$.

\begin{figure}
	\includegraphics[width=1\linewidth, angle=0]{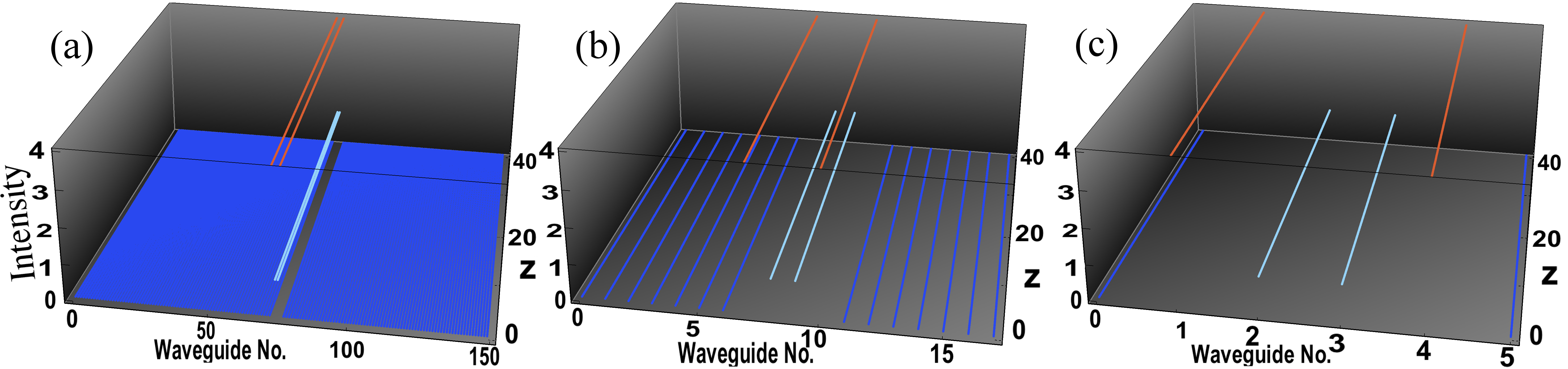}
	\caption{(Color online) Localization of the light at the exceptional point. (a) Beam dynamics in an array of 150 waveguides. We excited the mode associated with the flat band namely passive waveguide no. 73, loss waveguide no. 74 gain waveguide no. 75, passive waveguide no 76. Although the system is not Hermitian, the original excitation does not show any dynamics and the intensity remains constant. (b, c) The same as part (a) with 18 waveguides and two unit cells, namely 6 waveguides, respectively. The system size does not affect the mode associated with the flat band of infinite lattice.
		\label{fig3}}
\end{figure}
In general the dispersion relation of the PT symmetric Hamiltonian in equation \ref{eq1} does not have a closed form. However, at the exceptional point, using the expression of the $\gamma_{PT}$, one can find a closed form for the dispersion relation of the waveguide array
 \begin{equation}
 \label{eq2}
 \beta=\beta_0,-\frac{\beta_0}{2}\pm\sqrt{\frac{\beta_0^2}{4}+\frac{4kt^2 \cos ^2 \frac{q}{2}}{\beta_0}}.
 \end{equation}
Moreover, at the exceptional point the two bands coalesce which occurs at $q=\pm \cos ^{-1} (\frac{\beta_0^3-kt^2}{kt{^2}})$. This wavevector can be used to obtain the value of the gain and loss parameter for which we have the exceptional point
$\gamma_{PT}=\sqrt{\frac{2 k \beta_0 ^3}{t^2}+2 t^2 -3 \beta_0 ^2}$.
Further analysis shows that the propagation constant of the flat band is given by $\beta=\beta_0=t^2/k$ which can be used to obtain the $\gamma_{PT}$
\begin{equation}
\label{eq3}
\gamma_{PT}=t\sqrt{2-\frac{t^2}{k^2}}.
\end{equation}
Using equation (\ref{eq3}) and $H_{q=0}$ we can show that only for $|k|\geq|\frac{t}{\sqrt[4]{2}}$ the entire flat band is generated at the exceptional point.
 
 We can calculate the flat band mode in the spatial representation by taking the inverse Fourier of the eigenmode associated with the flat band in the momentum representation. We find that five sites need to be excited. More precisely, in unit cells $n^{th}$ and $n^{th}-1$, we need to excite one gain waveguide, one passive waveguide with amplitude one, $-\frac{k}{t}$, and one loss waveguide, one passive waveguide with amplitude $\frac{k^2}{k^2 -t^2-i k\gamma_{PT}}$, $-\frac{k}{t}\times\frac{k^2}{k^2 -t^2-i k\gamma_{PT}}$, respectively. 

To verify our analytical results, we perform numerical simulation for propagation distance $L=40$ coupling units. In our simulation in figure \ref{fig3}a, we consider 150 waveguides arranged according to figure \ref{fig1}. We excite the gain waveguide number 75 with amplitude $1$, waveguide number 76 with amplitude $-2$, waveguide number 74 with amplitude $\approx e^{0.72i}$, and waveguide 73 with amplitude $\approx -2 e^{0.72i}$. We see that initial excitation propagates without any diffraction and the light remains localized in the original excited waveguides. Furthermore, the initial intensity of excited waveguide remains constant and consequently the total norm associated with this excitation in the system is conserved.

In order to find the properties of the flat band generated by the exceptional point, we assumed that the PT symmetric waveguide array has infinite size. Therefore, we were able to find the dispersion relation of the lattice in equation \ref{eq2}. However, in the above numerical simulation we used only 150 waveguides which tells us that the compacton-like mode associated with the exceptional point might be robust against the changes in the system size. Hence to verify our conjecture, in figure \ref{fig3}b we perform numerical simulations for array of 18 waveguides which composed of 6 unit cells. We observe that the diffraction-less dynamics stays unchanged. In figure \ref{fig3}c we decrease the array size to the smallest number of unit cells that allow us to excite the dispersion-less mode namely 6 waveguides which composed of only two unit cells. Numerical simulation in figure \ref{fig3}c clearly shows that the compacton-like solution remains unaffected by changing the system size.

\begin{figure}
	\includegraphics[width=1\linewidth, angle=0]{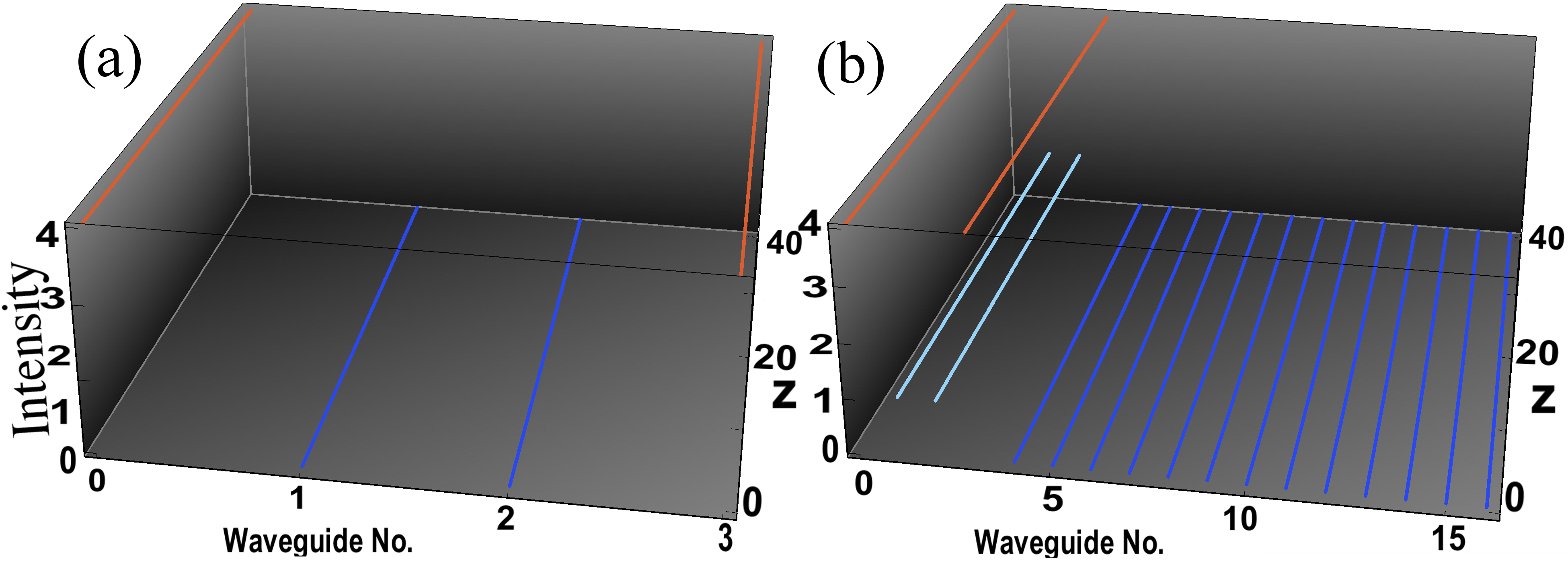}
	\caption{(Color online) Localization of the light at the edge of the lattice. (a) Four coupled waveguides composed of two passive waveguides, one loss and one gain waveguide forming \textit{B-C-A-B} type coupled waveguides. The excited mode propagates without any oscillation. Such a dynamic shows that the mode associated with the flat band of an infinite lattice at the exceptional point is an eigenmode of the coupled four waveguides. (b) Beam dynamics in six PT symmetric unit cells coupled from the left to a passive-loss (\textit{B-C}) dimer. Although the total system is not PT symmetric and have at least one complex propagation constant, the excited edge mode is not affected and remains at the edge with constant intensity.
		\label{fig4}}
\end{figure}

The robustness of the flat band localized mode at exceptional point vs system size helps us to explain how the entire flat band is generated at the exceptional point. In figure \ref{fig4}a in our numerical simulation, we coupled four waveguides, two passive waveguides at the edge and two active waveguides at the middle whereas one has gain and the other has loss (\textit{B-C-A-B}). This arrangement is the smallest waveguide number that accommodates the dynamic-less mode. We observe that the initial excitation propagates in the waveguides without any changes which indicates that such excitation is an eigenvector of the system. This clearly explains why the infinite lattice has a flat band at the exceptional point. By connecting these four waveguides through the dark states (loss waveguides with amplitude zero) a non-diffracting state with real propagation constant is generated. On the other hand such non-diffracting beam requires a flat band with no complex part. As the system originally does not have such band, the only option for the lattice is to find a flat band at the exceptional point. 

Furthermore, the robustness brings about the question of having the localized mode at the edge of the lattice. However, if we want to have the mode at the edge then at least one-unit cell should miss one waveguide (\textit{A} or \textit{C}) which makes the total system not to be PT symmetric anymore. Moreover, from Fig.\ref{fig4}a we infer that the localization can exists at the edge of a truncated lattice as depicted in figure \ref{fig4}b. In figure \ref{fig4}b, the simulation is performed in a waveguide array composed of six PT symmetric unit cells coupled from the left to a passive-loss (\textit{B-C}) dimer. So in this case the total system is not anymore PT symmetric and have at least one complex propagation constant. We clearly observe that the mode propagates at the edge of the lattice without any diffraction


Participation ratio $R=(\sum_{n}|\phi_n|^2)^2/\sum_{n}|\phi_n|^4$, where $\phi_n$ is the field in the $nth$ waveguide, is a measure of the degree of localization. Participation ratio takes value 1 for a state localized at one waveguide and N for completely delocalized state. In our case where 4 sites are populated the participation ratio is $R<3$ which indicates that we have an extremely localized state.

Our proposed structure can be demonstrated experimentally in numerous different systems \cite{27,34,35,36,37}. For instance in photonics, one can use the femtosecond direct writing method \cite{38} to realize array of passive-loss PT symmetric photonic coupled waveguide without using actual gain. In another approach, one can use the time reversed of the passive-loss PT waveguides, namely coupled laser cavities, to demonstrate non-Hermitian flat bands with the advantage of an exponential growth in non-diffracting super mode of coupled cavities. In the case of passive-loss (-gain) PT structure there are three sites in the unit cell of the periodic array. One with no net loss (gain), the second one with loss (gain) value $\gamma$ ($-\gamma$) and the third one with loss (gain) value $2\gamma$ ($-2\gamma$). The Hamiltonian of the system in the momentum representation in these cases is written as 
\begin{equation}
H_q=\left(\begin{array}{ccc}
0&-t&-k-ke^{-iq}\\
-t&-i(-)\gamma&-t\\
-k-ke^{iq}&-t&-2i(-)\gamma
\end{array}\right)
\end{equation}
If we make a transformation of the form $\psi_q=e^{-(-)\gamma z}\phi_q$ then the dynamics of the system in the new variable will be given by 
\begin{equation}
i\dot{\phi_q}={H_q} \phi_q ;{H_q}^\prime=\left(\begin{array}{ccc}
i(-)\gamma&-t&-k-ke^{-iq}\\
-t&0&-t\\
-k-ke^{iq}&-t&-i(-)\gamma
\end{array}\right)
\end{equation}
The above Hamiltonian ${H_q}^\prime$ has the same form as the one in Eq.(\ref{eq1}). Thus, our previous results are valid with the only difference that the mode amplitudes are multiplied by an exponential decay (growth) factor. 

In conclusion we demonstrated that altering non-Hermiticity in our PT symmetric system can change the flatness of bands. The ultimate robust flatness occurs at the exceptional point of the system where two bands combine to form an entire flat band embedded between the dispersive bands. The importance of our results is two folded, first it provides a controllable localization and BIC states and second it shows that localization can be used as a measure for the degree of non-Hermiticity.  Our proposal gives rise to new possibilities in imaging via gain and loss elements, quantum computing in the presence of complex entities, systems with intrinsic amplification or absorption mechanism such as coupled laser cavities and lossy metamaterials where Hermiticity is no longer valid, and long distance communication.


{\it Acknowledgments --} 
H.R gratefully acknowledge support from the UT system under the Valley STAR award.



\begin{thebibliography}{99}

\bibitem{1}	S. Fan, J. D. Joannopoulos, J. N. Winn, A. Devenyi, J. C. Chen, and R. D. Meade, J. Opt. Soc. Am. B 12, 1267 (1995).
\bibitem{2}	O. Painter, Science 284, 1819 (1999).
\bibitem{3}	J. Lydon, M. Serra-Garcia, and C. Daraio, Phys. Rev. Lett. 113, (2014).
\bibitem{t1} M. Kohmoto, B. Sutherland, and K. Iguchi, Phys. Rev. Lett. 58 2436 (1987).
\bibitem{t2}  W. Gellermann, M. Kohmoto, B. Sutherland , and P. C.Taylor, Phys. Rev. Lett. 72 633 (1994).
\bibitem{t3} M.H. Teimourpour, Journal of Optics 14 (3), 035501 (2012).
\bibitem{4}	M. Segev, Y. Silberberg, and D. N. Christodoulides, Nat. Photonics 7, 197 (2013).
\bibitem{5}	S. Mingaleev and Y. Kivshar, Phys. Rev. Lett. 86, 5474 (2001).
\bibitem{t4} M.H. Teimourpour, A. Rahman, K. Srinivasan, and  R. El-Ganainy, Phys. Rev. Applied 7, 014015 (2017).
\bibitem{6}	Y. Plotnik, O. Peleg, F. Dreisow, M. Heinrich, S. Nolte, A. Szameit, and M. Segev, Phys. Rev. Lett. 107, (2011).
\bibitem{7}	C. W. Hsu, B. Zhen, J. Lee, S.-L. Chua, S. G. Johnson, J. D. Joannopoulos, and M. Solja\v{c}i\'{c}, Nature 499, 188 (2013).
\bibitem{8}	S. Mukherjee, A. Spracklen, D. Choudhury, N. Goldman, P. \"{O}hberg, E. Andersson, and R. R. Thomson, Phys. Rev. Lett. 114, (2015).
\bibitem{9}	R. A. Vicencio, C. Cantillano, L. Morales-Inostroza, B. Real, C. Mej\'{i}a-Cort\'{e}s, S. Weimann, A. Szameit, and M. I. Molina, Phys. Rev. Lett. 114, (2015).
\bibitem{10}	V. Apaja, M. Hyrk\"{a}s, and M. Manninen, Phys. Rev. A 82, 41402 (2010).
\bibitem{11}	M. Hyrk\"{a}s, V. Apaja, and M. Manninen, Phys. Rev. A 87, 23614 (2013).
\bibitem{12}	M. C. Rechtsman, J. M. Zeuner, A. T\"{u}nnermann, S. Nolte, M. Segev, and A. Szameit, Nat. Photonics 7, 153 (2013).
\bibitem{13}	M. Biondi, E. P. L. van Nieuwenburg, G. Blatter, S. D. Huber, and S. Schmidt, Phys. Rev. Lett. 115, 143601 (2015).
\bibitem{14}	F. Guinea, M. I. Katsnelson, and A. K. Geim, Nat. Phys. 6, 30 (2009).
\bibitem{15}	S. Deng, A. Simon, and J. K\"{o}hler, J. Solid State Chem. 176, 412 (2003).
\bibitem{16}	M. Imada and M. Kohno, Phys. Rev. Lett. 84, 143 (2000).
\bibitem{17}	E. Tang, J.-W. Mei, and X.-G. Wen, Phys. Rev. Lett. 106, 236802 (2011).
\bibitem{18}	T. Neupert, L. Santos, C. Chamon, and C. Mudry, Phys. Rev. Lett. 106, (2011).
\bibitem{19}	S. Yang, Z.-C. Gu, K. Sun, and S. Das Sarma, Phys. Rev. B 86, 241112 (2012).
\bibitem{20}	S. A. Parameswaran, R. Roy, and S. L. Sondhi, Comptes Rendus Phys. 14, 816 (2013).
\bibitem{21}	T. Jacqmin, I. Carusotto, I. Sagnes, M. Abbarchi, D. D. Solnyshkov, G. Malpuech, E. Galopin, A. Lema\^{i}tre, J. Bloch, and A. Amo, Phys. Rev. Lett. 112, 116402 (2014).
\bibitem{22}	F. Baboux, L. Ge, T. Jacqmin, M. Biondi, E. Galopin, A. Lemaître, L. Le Gratiet, I. Sagnes, S. Schmidt, H. E. T\"{u}reci, A. Amo, and J. Bloch, Phys. Rev. Lett. 116, 66402 (2016).
\bibitem{23}	D. Leykam, S. Flach, O. Bahat-Treidel, and A. S. Desyatnikov, Phys. Rev. B 88, 224203 (2013).
\bibitem{konotop} A. V. Yulin and V. V. Konotop,  Opt. Lett. 38, 4880-4883 (2013).
\bibitem{24}	K. G. Makris, R. El-Ganainy, D. N. Christodoulides, and Z. H. Musslimani, Phys. Rev. Lett. 100, 103904 (2008).
\bibitem{25}	A. Szameit, M. C. Rechtsman, O. Bahat-Treidel, and M. Segev, Phys. Rev. A 84, (2011).
\bibitem{26}	B. Zhen, C. W. Hsu, Y. Igarashi, L. Lu, I. Kaminer, A. Pick, S.-L. Chua, J. D. Joannopoulos, and M. Solja\v{c}i\'{c}, Nature 525, 354 (2015).
\bibitem{27}	C. E. Ruter, K. G. Makris, R. El-Ganainy, D. N. Christodoulides, M. Segev, and D. Kip, Nat Phys 6, 192 (2010).
\bibitem{28}	H. Ramezani, H.-K. Li, Y. Wang, and X. Zhang, Phys. Rev. Lett. 113, (2014).
\bibitem{29}	H. Ramezani, Y. Wang, E. Yablonovitch, and X. Zhang, IEEE J. Sel. Top. Quantum Electron. 22, 115 (2016).
\bibitem{30}	C. M. Bender, Rep. Prog. Phys. 70, 947 (2007).
\bibitem{t5}	M.H. Teimourpour, R. El-Ganainy, A. Eisfeld, A. Szameit, and D. N.Christodoulide, Phys. Rev. A 90, 053817 (2014).
\bibitem{31}	D. L. Bergman, C. Wu, and L. Balents, Phys. Rev. B 78, 125104 (2008).
\bibitem{32}	H. Ramezani, D. N. Christodoulides, V. Kovanis, I. Vitebskiy, and T. Kottos, Phys. Rev. Lett. 109, (2012).
\bibitem{34}	A. Regensburger, C. Bersch, M.A. Miri, G. Onishchukov, D. N. Christodoulides, and U. Peschel, Nature 488, 167 (2012).
\bibitem{35}	J. Schindler, Z. Lin, J. M. Lee, H. Ramezani, F. M. Ellis, and T. Kottos, J. Phys. Math. Theor. 45, 444029 (2012).
\bibitem{36}	S. Weimann, M. Kremer, Y. Plotnik, Y. Lumer, S. Nolte, K. G. Makris, M. Segev, M. C. Rechtsman, and A. Szameit, Nat. Mater. 16, 433–438 (2017) doi:10.1038/nmat4811
\bibitem{37}	J. M. Zeuner, M. C. Rechtsman, Y. Plotnik, Y. Lumer, S. Nolte, M. S. Rudner, M. Segev, and A. Szameit, Phys. Rev. Lett. 115, (2015).
\bibitem{38}	A. Szameit and S. Nolte, J. Phys. B At. Mol. Opt. Phys. 43, 163001 (2010).

\end{thebibliography}
\end{document}